\documentclass[sigconf,nonacm]{acmart}

\renewcommand\footnotetextcopyrightpermission[1]{} 
\AtBeginDocument{%
  \providecommand\BibTeX{{%
    \normalfont B\kern-0.5em{\scshape i\kern-0.25em b}\kern-0.8em\TeX}}}

\setcopyright{none}
\copyrightyear{2019}
\acmYear{2019}
\acmDOI{10.1145/1122445.1122456}

\acmJournal{IMWUT}
\acmVolume{37}
\acmNumber{4}
\acmArticle{111}
\acmMonth{8}

\usepackage{tabularx}
\usepackage{booktabs}
\usepackage{hyperref}
\usepackage{graphics}
\usepackage{graphicx}
\usepackage{color,soul}
\usepackage{algorithm}
\usepackage[noend]{algpseudocode}
\usepackage{xcolor}   
\usepackage{pifont}   
\usepackage{subcaption}  
\usepackage{multirow}
\usepackage{wasysym} 
\usepackage{color, colortbl}
\usepackage{balance}
\usepackage{verbatim}
\usepackage{xspace}
\usepackage{balance}
\usepackage{amsmath, amsthm, amssymb, amsfonts}

\newcommand{\pioti}[1]{IoT Inspector}

\newcommand{\Figure}[1]{\textbf{Figure~\ref{fig:#1}}}

\renewcommand{\paragraph}[1]{\vspace{5pt}\noindent\textbf{#1: }}

\begin{document}

\title{Alexa, Who Am I Speaking To?}
\subtitle{Understanding Users' Ability to Identify Third-Party Apps on Amazon Alexa}

\author{David J. Major, Danny Yuxing Huang, Marshini Chetty$^*$, Nick Feamster$^*$}
\affiliation{Princeton University, University of Chicago$^*$ \\ ~ \\ Project Website: \url{https://sites.google.com/view/alexawhoamispeakingto/}}
\authorsaddresses{}

\renewcommand{\shortauthors}{}

\begin{abstract}

Many Internet of Things (IoT) devices have voice user interfaces (VUIs).  One
of the most popular VUIs is Amazon's Alexa, which 
supports more than 47,000 third-party applications
(``skills'').  We
study how Alexa's integration of these skills may
confuse users. Our survey
of 237 participants found that users do not understand that skills
are often operated by third parties, that they often confuse third-party
skills with native Alexa functions, and that they are unaware of the
functions that the native Alexa system supports.  Surprisingly, users who
interact with Alexa more frequently are more likely to
conclude that a third-party skill is native Alexa functionality.  The
potential for misunderstanding creates new security and privacy risks: attackers can develop third-party
skills that operate without users' knowledge or masquerade as native
Alexa functions.  To mitigate this threat, we make design recommendations to help users distinguish
native and third-party skills.

\end{abstract}



\settopmatter{printfolios=true}

\maketitle

\section{Introduction}

Voice User Interfaces (VUIs) are a common way to interact with an increasing
number of smart-home Internet of Things (IoT) devices. On Amazon Echo (and
related devices, such as the Echo Dot), the most popular type of voice
assistant~\cite{SADeviceRanking1}, a VUI known as the Alexa voice service
(``Alexa'') enables users to use only voice to interact with the device's wide
range of functions. These so-called \textit{skills} (analogous to applications
or ``apps'' on mobile devices) range from setting an alarm clock, telling
jokes, to sending money.  Although some of the skills ship with Alexa by default,
\textit{native} to the Alexa platform, there are more than 47,000 skills that
are developed by third parties, which users can invoke to add new functions to
their Echo devices~\cite{skill-store}.
Amazon has made certain decisions concerning the integration of these skills
that sometimes make it difficult to determine the agent responsible for
implementing the skill (i.e., whether it is Amazon or some other third party).

In this paper, we seek to understand the extent to which these design
decisions make it difficult for users to determine whether they are
interacting with native Amazon functionality or with arbitrary third parties.
To do so, we conducted a survey between March and May 2019 with 237 new and existing users of Alexa,
including 103 undergraduates, and 134 Amazon Mechanical Turk workers. Most
participants were unaware that Alexa skills can be (and often are) developed
by third-party developers. Our work is distinct from previous work on VUIs in
this regard, which has focused more generally on how users interact with VUIs~\cite{8560072,Lopatovska:2018:PAA:3176349.3176868,Purington:2017:AMN:3027063.3053246,Porcheron:2018:VIE:3173574.3174214,Sciuto:2018:HAW:3196709.3196772,doi:10.1080/10447318.2014.986642},
but not specifically on their interactions with third-party skills, or their
ability to distinguish third-party skills from native ones.

In each survey, we present each participant with video and audio clips of
interactions with three types of skills in the lab---(1)~Alexa's native
skills, (2)~publicly available third-party skills, and (3)~malicious skills
that we developed and which are not publicly available---without revealing to
the participant which exactly which type of skill is being presented. The
participants' reactions to these clips suggest that, even after learning that
some Alexa skills are developed by third parties, the participants were often
unable to differentiate between native and third-party skills or between
native skills and malicious third-party skills, and that the participants did
not understand what functionalities are possible through native or third-party
skills.

We discovered that Alexa users are unable to differentiate
native skills from third-party skills. Much to our surprise, we found that
users who have {\em more} familiarity and experience with Alexa are in fact
more likely to mistakenly assume that a third-party skill is in fact native
Amazon functionality.
We also found that some participants were
unaware that skills could be developed by third parties, that most
participants failed to distinguish native and third-party skills and voice
messages, and that they often did not understand what functions or voice
commands were available on Alexa. As a result, a user may confuse a skill,
whether malicious or benign, with a native skill or another benign skill, thus
potentially exposing sensitive information to an unintended third party.

Our findings suggest that participants' misunderstandings of Alexa skills are
rooted in two design decisions that run counter to Norman's design principles~\cite{DesignOfEverydayThings}.
First, Alexa's responses from native functions sound the same as third-party
skills. Presumably this design decision was made to create a seamless user
experience, but it can also cause confusion concerning whether the skill is
native to the Echo or whether the skill constitutes a third-party application.
In contrast to graphical user interfaces (GUIs) on computers and mobile
devices, which typically offer visual cues that aim to help users distinguish
between applications, VUIs can have more difficulty providing direct cues to
users.  For example, the Echo has a colored light that indicates the state of
the device, but it does not offer any information concerning the skill the
user is interacting with~\cite{alexa-light}. As such, a user may not be aware
of whether he or she is interacting with a native skill or third-party skill, or
specifically which third-party skill has been invoked. This confusion could
ultimately lead to security and privacy risks. For example, past research has
shown that a malicious third-party developer could develop a new skill that
mimics the behavior of an existing benign skill in an attempt to trick users
into revealing sensitive information~\cite{Kumar:2018:SSA:3277203.3277207,
DBLP:journals/corr/abs-1805-01525}.

Second, Alexa has effectively infinite voice commands, which makes it
difficult for users to understand what and how different skills being invoked
and what voice command might potentially invoke a skill.  In particular, a
user could invoke the same skill with different commands; for example, ``set
an alarm at 8 am'' and ``wake me up at 8 am'' achieve the same functionality
on Alexa. Each of the 47,000 third-party skills may define arbitrary voice
commands. For example, there are at least 200 joke-related
skills~\cite{joke-skills}, each with its own voice commands, ranging from
``Open knock knock''~\cite{joke-knock} to ``Ask daily jokes to give me a
joke~\cite{joke-daily}.'' Because a typical user is unlikely to known or
remember all available commands, users can find it difficult to know what
functionality is and is not supported by Alexa. The inability to distinguish
native and third-party skills has potential security and privacy implications:
A malicious third-party developer could take advantage of this design, create
a skill that responds to user commands that might otherwise be unrecognizable
by Alexa, and trick the user into using the malicious skill.

These findings suggest clear directions for design improvements, including
better ways to users to distinguish native from third-party contexts, as
well as consistent design standards and interaction modes for third-party
skills that run on the Alexa platform.

\section{Background: Alexa Skills}

Recent years have seen a proliferation of voice-enabled IoT devices, ranging
from phones, voice assistants, microwaves, to car navigation systems.  This
paper focuses on one specific type of voice-enabled device can host
third-party applications. In this sector, Amazon is the dominant player with
61\% market share across its Alexa-enabled Echo devices (Google has the second
highest with 17\%) \cite{SADeviceRanking1}. To further spread Alexa, Amazon
has built the Alexa Voice Service which can configure other smart devices (not
made by Amazon) to run Alexa software \cite{AVS1}. Thus, Alexa can be seen as
the clear leader in the field and a useful case study for understanding how
users interact with VUIs for virtual assistants.  We provide an overview of
Alexa's skill ecosystem and a description on how users invoke and interact
with skills.

\subsection{Native and Third-Party}

Alexa supports two types of skills: (1)~native skills and (2)~third-party skills. Native
skills come built-in by Amazon and thus only involve code and dialog developed
by Amazon. For example, users can ask for the time, set an alarm, or play music
from Amazon Music. As Amazon is the sole developer for these skills, we assume
that all information collected from users flows only to Amazon.

To support a broader range of functions, Amazon allows third-party developers to
build skills for Alexa using the Alexa Skills Kit. Using the skills kit,
developers can configure Alexa to communicate with their own services, create
custom Alexa responses, and run custom code on their own servers~\cite{AVS2}.
Third-party developers have built at least 47,000 skills, including a wide
variety of functions such as playing podcasts, locking doors, checking credit
card balances, and telling jokes, that are publicly available on the Amazon
Skill Store~\cite{skill-store}. Since the code of these skills could be on third-party servers, we
assume that some of the information collected from users may flow to the
third-party developers (in addition to Amazon).

\subsection{Invoking Skills}

Whether a skill is native or third-party, a user can invoke (i.e., verbally
enable) it by saying the corresponding \textit{invocation phrases}. These
phrases follow the form of ``Open <invocation name> for <optional action>''
where the invocation name is often the name of the skill. Examples include
``Alexa, open Jeopardy'' (i.e., a game shown in the US) and ``Alexa, ask Daily Horoscopes about Taurus.''

However, Alexa allows some flexibility in invoking skills. For some native
skills such as the alarm clock, a user can invoke it via either ``Alexa, set an
alarm for 8 am'' or ``Alexa, wake me up at 8 am.'' For third-party skills, users
replace ``Open''  with one of 13 words such as ``play'', ``open'', and
``launch''. If none of these phrases are present, Alexa automatically parses the
user's statement for an invocation name and responds with the corresponding
skill~\cite{AVSSkill}. However, invocation names do not appear to be unique, as
we have found skills with the same invocation names. It is unclear how Alexa
chooses which skill to invoke given two skills with the same invocation name.

\subsection{Interacting with Skills}
Once a user invokes a skill, Alexa enters what we call the skill's
\textit{context}. At the time of writing, Alexa does not visually (through the
device lights) or verbally confirm which context a user is in; in fact, Alexa's
voice sounds exactly the same. Once Alexa is in a skill's context, Alexa accepts
only voice commands predefined by that skill, along with phrases such as
``cancel'' or ``quit'' that allow users to leave the skill's context and invoke
a different skill. A user cannot invoke a second skill until the user leaves the
first skill's context.

\section{Related Work}

In this section, we provide a literature review of VUIs, focusing on how to design VUIs, how humans interact with VUIs, and security/privacy concerns regarding VUIs.

\subsection{Designing Voice User Interfaces}

Design patterns for graphical user interfaces are a well-established
field~\cite{Laurel:1990:AHI:575201}, but paradigms for VUI design are scarce to
our knowledge, presumably because voice assistants and other voice-enabled
technologies have only taken off in recent years. One example of literature in
VUI designs is Cathy Pearl's \textit{Designing Voice User Interfaces: Principles
of Conversational Experiences}~\cite{DesigningVUICathyPearl}, which covers
design principles such as always confirming users' voice input or handling
ambiguous voice commands. However, the authors assumes that only the first party
(i.e., the device manufacturer) engages in conversation with users without
considering third-party capabilities such as skills. Similarly, L{\'o}pez et
al.~\cite{10.1007/978-3-319-60366-7_23} evaluated the usability of popular VUIs
such as Alexa and Apple Siri in terms of correctness of responses and how
natural the responses sound; again, this work did not consider third-party
functionalities. In fact, we are unaware of any literature in VUI design that
incorporates skills, and we are among the first to discuss skill-enabled VUI
design in the research community.

Despite the apparent lack of literature, there are general design principles
that could apply to our case. Don Norman's \textit{Design of Everyday
Things}~\cite{DesignOfEverydayThings} introduces seven fundamental principles
of design, three of which are especially relevant to this study of VUIs: (1)
Discoverability, which, when applied to skills, suggests that Alexa should let
users know what voice commands are available; (2) Feedback, which suggests
that Alexa should inform users of which skills they are currently interacting
with; and (3) Conceptual Model, which would require Alexa to help users
understand that skills are developed by third parties. As we will show in the
survey results, Alexa's design appear inconsistent with these principles, thus
exposing users to security and privacy risks.
We leave for future work to
evaluate Alexa's design against the remaining four design principles:
affordances, signifiers, mappings, and constraints.

\subsection{Human Interactions with Voice User Interfaces}

A large body of work studies how humans interact with VUIs and what kind of relationship is developed as a result. For instance, researchers found that some users personified their VUI devices and treated the devices with emotion as if the devices were family members or friends~\cite{8560072,Lopatovska:2018:PAA:3176349.3176868,Purington:2017:AMN:3027063.3053246}. Past work has also found that interactions with VUIs were integrated with activities or conversations involving the entire household, including children~\cite{Porcheron:2018:VIE:3173574.3174214,Sciuto:2018:HAW:3196709.3196772}. However, some researchers identified privacy concerns for VUIs in the public space, resulting in greater user caution when users transmitted sensitive information than non-sensitive information~\cite{doi:10.1080/10447318.2014.986642}. In this paper, we also study how users interact with a VUI (i.e., Alexa), but we specifically focus on how users could be confused by Alexa's design and how users might leak sensitive information due to this confusion.

\subsection{Security and Privacy Risks}

Users face multiple security and privacy risks that originate from a number of
actors. First, manufacturers of voice assistants, i.e., the first
parties, may collect potentially sensitive recordings of users without the
users' knowledge, for instance, through the always-on microphones. This design may lead to
accidentally recording sensitive conversations and sharing the data with the
manufacturers~\cite{malkin2019privacy, 8048642}. In addition to manufacturers, third-party skills (or ``actions'' on Google Home)
could also present security and privacy risks to users.
In particular, a third-party malicious skill could effectively phish a user by
pretending to be another benign skill. As demonstrated in a proof of concept
by Kumar et al.~\cite{kumar2018skill} and Zhang et
al.~\cite{zhang2019dangerous}, a malicious skill could use an invocation name
that sounds similar to a benign skill, such as ``Capital One'' (legitimate
banking skill) and ``Capital Won'' (malicious skill). 

At the time of writing,
neither Amazon Alexa nor Google Home provides users with audio or visual clues
regarding which skill the user is interacting with. This design decision may
result in users invoking the wrong skills and providing sensitive information
to malicious third parties.
While both of these papers focused on skills with similar-sounding names,
we study how users could be confused by skills \textit{in general} (including those with similar-sounding names). Also, whereas these papers demonstrated attackers'
capabilities, we empirically show, from the user's perspective, that a user
could fall prey to such malicious skills.

\section{Survey Method}

To understand how users conceptualize and interact with Alexa and its skills, we
conducted surveys of both Alexa owners and non-owners in two populations: 103 undergraduate and graduate university students (``University Survey'') and
134 participants through Amazon Mechanical Turk (``MTurk Survey''). Having both surveys enabled us to survey a wide swath of participants~\cite{redmiles2019well}. We tested both owners and non-owners to better understand whether previously owning or using an Alexa affected a participant's familiarity with the device and how skills operate.
Both surveys were approved by our university's Institutional Review Board (IRB).

\subsection{Recruitment}

\textbf{Recruitment method:} We conducted the University Survey between March 27 and April 10, 2019.
We recruited 103 U.S. university student participants by email through our
university's survey research center (SRC). The SRC randomly selected students
and emailed them a link to the survey hosted on Qualtrics. We incentivized
participation by awarding approximately 1/10 participants with an Amazon Echo
device. We did not require Alexa ownership or experience in the recruitment criteria, although participants who decided to take the survey were presumably aware of or interested in Alexa.

Based on our initial findings, we expanded the survey (i.e., asking participants how often they used Alexa) and conducted the MTurk Survey between April 19 and May 9, 2019.
Through Amazon Mechanical Turk (MTurk), we recruited 134 English-speaking
participants with at least a bachelor's degree. Participants were paid
at least minimum wage for the ten-minute survey.
To ensure quality responses, we shared the survey only with MTurk users with
approval ratings over 95\% and who were MTurk ``masters'', a special designation
Amazon gives only to the top-performing MTurk users. Again, Alexa ownership or experience was not a part of the recruitment criteria.

\textbf{Removing participants:} To exclude low-quality responses and their participants, we added multiple attention checks throughout Survey Sections 3 and 4 (i.e., after we described key concepts such as skills in Survey Section 2), such as ``who build native Alexa skills'' and ``who build third-party Alexa skills,'' with ``Amazon'' and ``non-Amazon'' as the choices. In doing so, we hoped to remove participants who failed to understand the concept of third-party skills. In a further effort to filter out low-quality participants, we also prevented re-try attempts and added a time-limit to the survey.

\textbf{Characteristics of participants:}
Of the participants reached through our university's
SRC, all were current undergraduate or graduate students. While participants
came from a wide range of specific majors, about 40.4\% were in an
engineering-related subject. 51.0\% identified as male, 47.0\% as female, and 2.0\% as other. 19.6\%
of participants own an Alexa. We did not ask how often the participants used their Alexa devices.

For the MTurk survey, all participants
have a bachelor's degree (as verified by Mechanical Turk), a 95\%
approval rating or higher, and Mechanical Turk ``Masters'' status. 51.9\%
identified as male and 48.1\% as female. 75.2\% of participants own an Alexa. Of
those that own an Alexa, 34.0\% have owned it for less than a year, 40.0\% for 1-2
years, and 26.0\% for over 2 years. A majority (56.1\%) use it several times a
day while 9.2\% use it less than once a week, 7.1\% once a day, and 27.6\%
several times a week.

\textbf{Limitations:}
Because our study is limited to current university students and users of Amazon
Mechanical Turk, our overall subject population is likely more
tech-savvy than the general population. For example, 86\% of MTurk participants
owned either an Alexa or some other smart device (smart device being defined as
any Internet-connected device other than a computer or phone). Nevertheless, we
believe that our results are generally applicable; if anything, the general
population that is less tech-savvy would highlight a general lack of awareness
about Alexa skills than our survey results already do with the tech-savvy
population.

\subsection{Survey Questions}\label{sec:survquestions}

The general goals of the survey were to understand what are participants'
privacy expectations with regard to native and third-party skills on Alexa,
whether participants can differentiate between native and third-party
functionality, and whether participants know what voice commands skills accept.
To this end, we included the following four sections in the Qualtrics survey.

\subsubsection{Survey Section 1: Pre-definition Questions about Alexa}

Through this survey section, we aimed to understand users' privacy expectations of skills
\textit{before} we defined the terms native/third-party skills for them. Per Norman's
design principles~\cite{DesignOfEverydayThings}, our goal was to check whether
users' Conceptual Model of third-party skills was consistent with the reality.

To this end, we asked the following questions. We first asked a general
true/false question, ``Everything Alexa says is programmed by Amazon,'' and we
counted the number of respondents with each answer ``yes'' or ``no'' (correct answer).

Additionally, we produced and presented the following videos in order, in which
a member of the research team engaged in a conversation with Alexa in the lab.
We then asked users where they thought the data from the conversation was sent:
to ``Only Amazon'', ``Only Third Parties'', or ``Both''. Interested readers can
view all our videos (including those in later survey sections) at our anonymized project webpage,
\url{https://sites.google.com/view/alexawhoamispeakingto/}.

\begin{itemize}
    \itemsep=-1pt
\item \textbf{Video 1A: Add Rubber Ball to Shopping Cart} \\
 A user (i.e., a member of the research team) asks Alexa to add a rubber ball to his cart, and Alexa responds, ``Ok, I've added a choice for rubber ball to David's Amazon Cart.'' This is an actual interaction that occurs when a user asks Alexa to add an item to the cart. In the survey, we asked each participant, ``Immediately as result of the following conversation, what parties do you think know David added a rubber ball to his Amazon cart?'' The correct response is ``Only Amazon.''

\item \textbf{Video 1B: Bedtime Story} \\
The following conversation occurs: \\
\textit{User}: ``Alexa, open `I'm going to bed'.'' \\
\textit{Alexa}: ``Time for a bedtime story! First, what's your name?'' \\
\textit{User}: ``Benji.'' \\
\textit{Alexa}: ``Ok, Benji! Here's your story.'' \\
This skill is an example third-party bedtime story skill we built (which we did not publicly release). In the survey, we asked each participant, ``Immediately as result of the following conversation, what parties do you think know your name is `Benji'?'' The correct response is ``Both.''

\end{itemize}

After showing Video 1B, we asked each participant to provide a free-text open-ended response to explain their rationale for their answer.

\subsubsection{Survey Section 2: Defining Key Alexa Concepts/Terms}

In this survey section, we briefly described to participants what an Alexa skill
is and what native and third-party skills mean. The goal is for us to ensure, to
our best effort, that the participants understood these concepts in later
sections, as we would test whether participants could distinguish between native
and third-party skills and capabilities.



\subsubsection{Survey Section 3: Differentiating Native and Third-Party Skills}

To test whether participants could differentiate between native third-party
skills -- effectively whether Alexa was able to offer Feedback (per
Norman's design principles~\cite{DesignOfEverydayThings}) on which skills a user
was interacting with -- we embedded five video clips and five audio clips in
this section of the Qualtrics survey and asked for the participants' response.
Similar to Videos 1A and 1B, we produced the following clips ourselves and
presented them to the participants in order.



The video clips show a member of the research team interacting with a native or
third-party skill. After we showed each clip, we asked participants whether the participant had interacted with a native or third-party Alexa skill.

\begin{itemize}
    \itemsep=-1pt

\item \textbf{Video 3A: Tell a Joke (native).}
    A user asks Alexa for a joke and Alexa responds with a joke.

\item \textbf{Video 3B: Jeopardy (third-party).}
    A user asks Alexa to play Jeopardy (a US game show) and the game begins with the voice of Alex Trebek (Jeopardy's host).

\item \textbf{Video 3C: Baseball Scores (third-party).}
    A user asks Alexa about the Astros (a baseball team) and Alexa responds with
    the latest scores.

\item \textbf{Video 3D: Rain Sounds (third-party).}
    A user asks Alexa to play rain sounds and Alexa responds with the sound.

\item \textbf{Video 3E: Parental Controls (third-party).}
    A user asks Alexa to enable parental controls and Alexa responds confirming
    the user would like to do that. While Videos 3B through 3D feature real
    skills available on the Amazon Skill Store, the parental control skill in this
    video is not public; in fact, we developed this skill ourselves using Alexa
    Skill Kit and made it available only to the Amazon Echo in our lab. We
    designed this skill to sound as if it could configure parental controls on
    Alexa, although in reality parental controls cannot be configured verbally
    with Alexa.

\end{itemize}

We also showed audio messages that we recorded from a native skill or third-party skill. The third-party skill could be from the Skill Store~\cite{skill-store}, or it could be developed by us and not released publicly. We asked each participant to respond whether the message was a real system message (i.e., native skill) or a fake one (i.e., third-party skill masquerading as a native skill).


\begin{itemize}
    \itemsep=-1pt

    \item \textbf{Audio 3A: Wi-Fi (fake).} ``You seem to be disconnected from
    Wi-Fi. Please hold down the circle button in order to reconnect.'' Similar to
    the skill in Video 3E, we developed a private skill that hard-coded the
    message above. Following the instructions would initiate a hard reset of the
    device.

    \item \textbf{Audio 3B: Problem with Response (real).} ``There was a
    problem with the requested skill's response.'' Alexa generates this verbal
    message when a third-party skill's response is not configured correctly.

    \item \textbf{Audio 3C: Link (fake).} ``Sorry, something is wrong with
    this device. Please restart or go to amazon.com/alexa for more
    information.'' Again, we developed this private skill ourselves. A malicious
    third-party skill could say this message (e.g., in the middle of other
    activities of the skill, thus giving the illusion that this is a
    system-generated message) , replacing the URL with that of a phishing
    website to obtain sensitive user information.

    \item \textbf{Audio 3D: Sorry (real).} ``Sorry, I'm not sure about that.''
    Alexa generates this message when it cannot understand the user's voice
    commands.

    \item \textbf{Audio 3E: Amazon Account (fake).} ``Sorry, before using this
    device you need to connect your Amazon account.'' We developed this private
    skill ourselves.

\end{itemize}

\subsubsection{Survey Section 4: Voice Commands that Alexa Understands}

Finally, we aimed to test whether Alexa offers users Discoverability,
per Norman's design principles~\cite{DesignOfEverydayThings}, or whether users
know what voice commands can be understood by Alexa.

In particular, we asked participants whether the following invocation phrases could
open skills on Alexa: ``Open Grubhub,'' ``What's the NY Times report,'' ``Find my iPhone,'' ``Quit,'' ``Please go away,'' and ``There's a bug over there.'' With the exception of ``Quit'' (which lets users leave a particular skill), all these phrases can open actual Alexa skills on the Skill Store or those we developed in private (e.g., ``Please go away'' and ``There's a bug over there.'').

We also asked whether certain actions can be accomplished with Alexa
verbally: changing device volume, muting device, checking WiFi connection, changing Amazon password, ordering items on Amazon, turning off device, and turning on/off parental controls. At the time of writing, the only actions that Alexa can accomplish are changing device volume and ordering Amazon items. These questions
are relevant,  as participants' expectations of what can be done on Alexa and
what invokes third parties on Alexa can influence their ability to differentiate
between native and third-party skills.

\begin{figure}[t]
\centering
\includegraphics[width=\columnwidth]{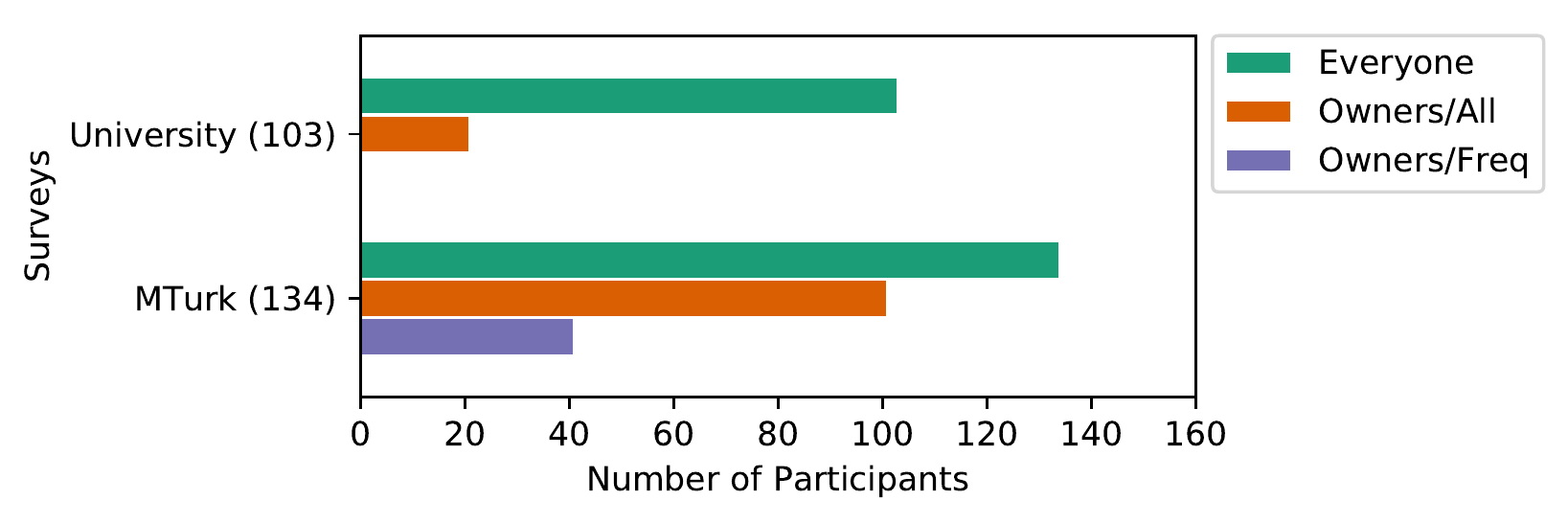}
\caption{Number of participants. The numbers in parentheses indicate the sample size.}
\label{fig:sample_features}
\end{figure}

\subsection{Data Analysis}

\textbf{Preparing data for analysis:} We downloaded the survey responses from Qualtrics as CSV files and analyzed the data in Python Pandas. We removed 3 University participants and 48 MTurk participants for failing the attention checks.

\textbf{Labeling participant groups:} As will be discussed in the Findings section, we analyzed the responses in terms of different levels of familiarity and experience with Alexa. To facilitate this analysis, we created three participant groups: (i) ``Everyone, '' which refers to all 237 participants; ``Owners/All,'' which is a subset of ``Everyone'' that refers to those that own Alexa devices, including 21 and 101 participants in the University and MTurk surveys respectively; and ``Owners/Freq,'' which is a subset of ``Owners/All'' that includes owners of Alexa that had owned the device for at least a year and indicated usage ``multiple times a day or more.'' We used these labels to denote users with potentially increasing levels of familiarity and experience with Alexa. Since we did not ask how often University participants used Alexa, all 41 ``Owners/Freq'' participants were from the MTurk survey. We provide a summary in \Figure{sample_features}.

\textbf{Coding free-text responses:} For each free-text open-ended survey question, one member of the research team coded all responses using qualitative techniques~\cite{saldana_coding_2013}. Example codes tagged phenomena of interest related to a participant's mental model of Alexa -- for instance, whether Amazon alone handled the interaction, or whether a third party was involved. Another member of the team then individually reviewed the codes and we discussed final themes as a research team. For both free-text survey questions, the second team member was able to validate all the codes/responses without disagreement.

\section{Results}\label{sec:results}

Our survey results yield three major themes:

\begin{enumerate}
    \itemsep=-1pt
    \item Many participants are unaware that Alexa skills can be developed by third parties.
    \item Even when informed that Alexa skills can be developed by third
        parties, most participants could not differentiate between native
        Alexa functionality and third-party skills, nor between real and fake
        Alexa system messages. Interestingly, frequent users were even less
        able to distinguish native from third-party skills.
    \item Alexa users often do not understand the standards of how the Alexa
        system functions nor what is possible and not possible on Alexa.
\end{enumerate}

\subsection{Finding 1: Participants are unaware that skills are developed by third parties}\label{sec:mental}

Our results showed that participants' Conceptual
Models~\cite{DesignOfEverydayThings} of who develops skills and
who could see the users' data are inconsistent with the reality, where third
parties can build skills and thereby have access to user behavioral data~\cite{AlexaSaveData}.

\begin{figure}[t]
\centering
\includegraphics[width=\columnwidth]{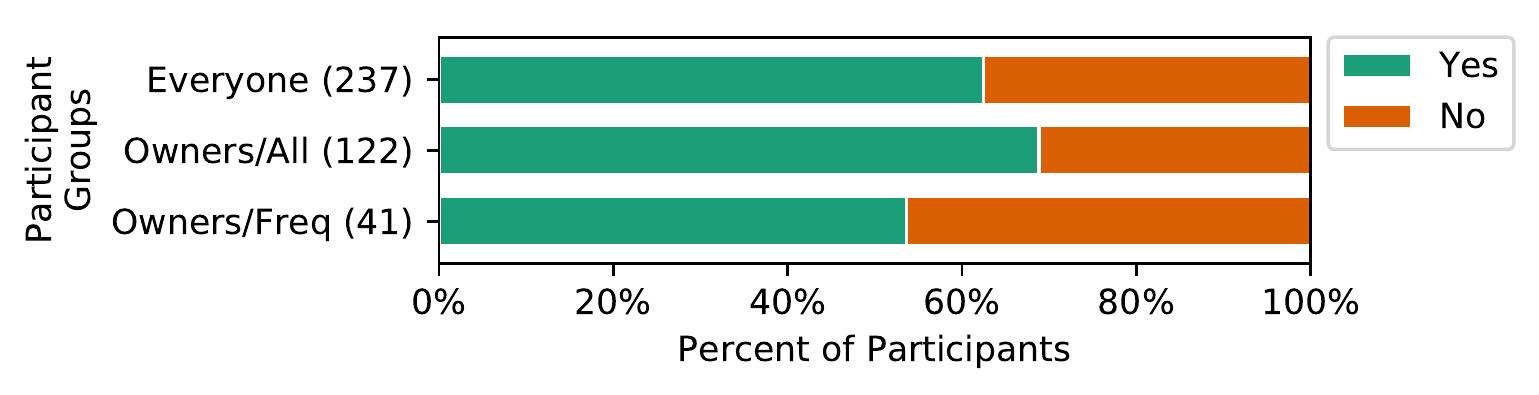}
\caption{Responses to the question, ``Everything Alexa says is programmed by Amazon.'' Correct answer: ``no.''}
\label{fig:programmed_by_amazon}
\end{figure}

\subsubsection{Some participants assume all Alexa contents/capabilities are handled by first party}

The participants' Conceptual Model of the device, particularly with regard to
who builds skills, runs counter to reality. As shown in
\Figure{programmed_by_amazon}, when asked whether ``everything Alexa says is
programmed by Amazon'' in Survey Section 1 (with ``no'' as the correct answer),
62.4\% of all participants (``Everyone'') thought the statement was true. In
particular, 68.9\% of ``Owners/All'' answered ``yes,'' which suggests that
familiarity with Alexa may not always correspond to a more accurate Conceptual
Model.

\subsubsection{Some participants are unaware that third-party skills could collect data}

\begin{figure}[t]
\centering
\includegraphics[width=\columnwidth]{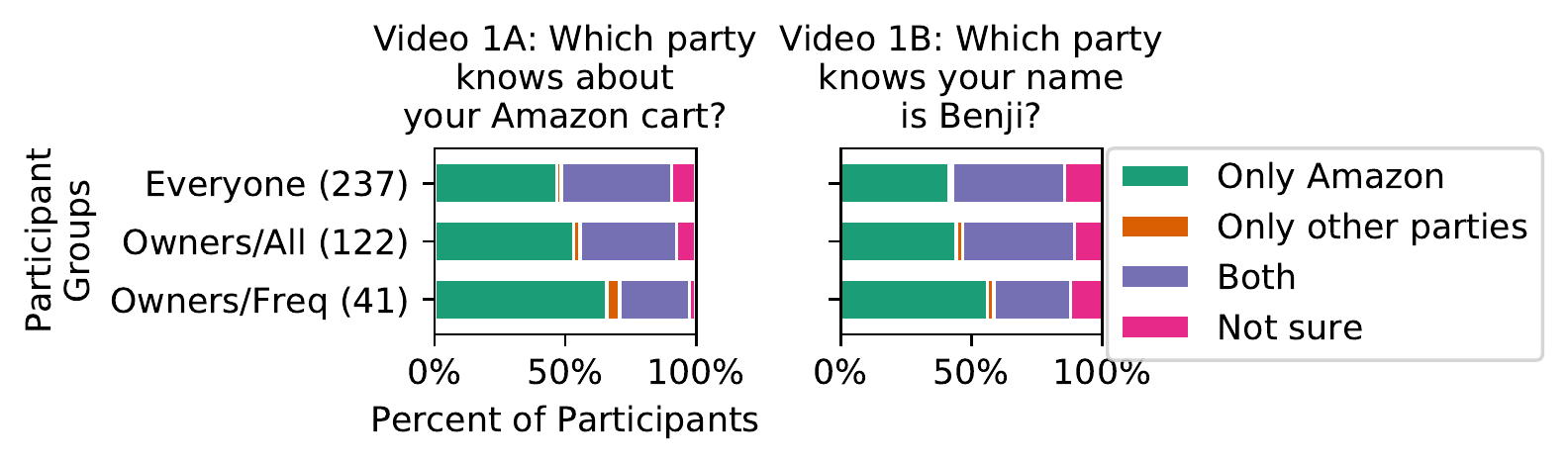}
\caption{Responses two videos. Video 1A: What parties do you think know David added a rubber ball
    to his Amazon cart? Correct answer: ``Only Amazon.'' Video 1B: What parties do you think know your name is Benji? Correct answer: ``Both.''}
\label{fig:video_1a_1b}
\end{figure}

In Survey Section 1, some participants were unaware that third parties could collect user data through Alexa skills. \Figure{video_1a_1b} shows participants' responses to Videos 1A and 1B that were meant to gauge whether the participant understood that Amazon had third-party skills, and that the third parties had access to user responses. In particular, 46.8\% of ``Everyone'' understood that only Amazon had access to the cart information (Video 1A), and this percentage increased as the level of familiarity and experience increased; in fact, some 65.9\% of ``Owners/Freq'' answered correctly. For Video 1B, however, the more experienced participants were associated with lower rate of correctness; for example, 56.1\% of ``Owners/Freq'' incorrectly believed only Amazon knew the name was Benji, compared with only 41.4\% of ``Everyone.'' Again, familiarity with Alexa may not always correspond to a more accurate Conceptual
Model.

\begin{figure}[t]
\centering
\includegraphics[width=\columnwidth]{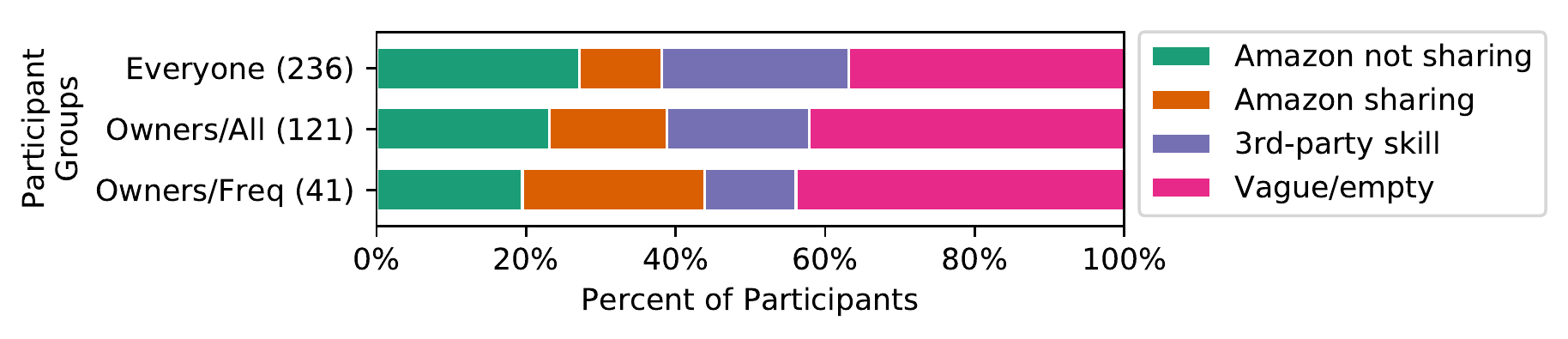}
\caption{User explanation (coded) for their responses to Video 1B.}
\label{fig:video_1b_reason}
\end{figure}

In an open-ended free-text question after Video 1B, we asked why each participant answered in a certain way. Three themes emerged from the responses: (i) Amazon originally having the user data and subsequently sharing it with third parties (``Amazon sharing''); (ii) Amazon originally having the data but not sharing it with third parties (``Amazon not sharing''); and (iii) understanding that a third-party skill could directly have access to the data (``3rd-party skill''). For empty responses or vague responses, we used the code ``Vague/empty.''

We present a distribution of these codes in \Figure{video_1b_reason}, which suggests that only 25.0\% of ``Everyone'' understood that skills had direct access to the data, rather than relying on Amazon to share the data. This percentage decreases as the level of familiarity and experience with Alexa increases; in fact, only 12.2\% of ``Owners/Freq'' made the same choice. An example of a response showing this understanding included \textit{S1R13}, \emph{``Data is shared by the third party developer of the app''} and  another participant, \textit{S2R8}, who wrote: \emph{``I think it is a skill developed by another party, and they will have access to this data''}.

For participants that were not aware of the skills, many believed that their interactions with Alexa were strictly with Amazon. Overall, 27.1\% of ``Everyone'' were coded ``Amazon not sharing.'' For instance, \textit{S1R30} wrote: \emph{``As far as I'm aware, Amazon doesn't sell any data to other companies, it only uses it privately (I could be wrong but I think this is true).''} Similarly, \textit{S2R23} responded: \emph{``Alexa is connected to Amazon and I think most info is stored and shared only with Amazon.''} In contrast, 11.0\% of ``Everyone'' believed Amazon did share data with third-parties (as opposed to skill developers having direct access to the data). For instance, \textit{S1R5} wrote: ``There have been enough reports of information sharing across 'The Internet of Things' for me to presume that any information given to a smart device, especially one belonging to the Amazon company, is shared with other parties and services.'' and \textit{S2R24} wrote: ``I don't trust anyone to not sell or share data. They all do it.''

In summary, most of these responses (38.1\% of ``Everyone'') centered around whether or not Amazon shared the data rather than the interaction being with a third party itself. While participants' opinions on data sharing is irrelevant to this paper, their responses shed light on their Conceptual Model of Alexa. For a majority of participants, an Alexa user interacts directly with Amazon alone, and only Amazon possesses data from the exchange as a direct consequence. This Conceptual Model contrasts greatly with the reality, where a skill can be built by any developer and anything a user says in such an interaction can go directly to the developer.

\begin{figure*}[t]
\centering
\includegraphics[width=\textwidth]{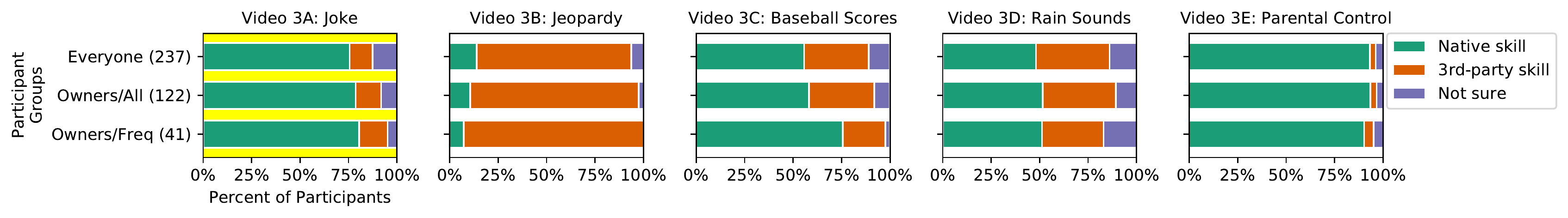}
\caption{Differentiating between native and third-party skills. Only Video 3A (highlighted) shows a native skill.}
\label{fig:video_3a_3e}
\end{figure*}

\begin{figure*}[t]
\centering
\includegraphics[width=\textwidth]{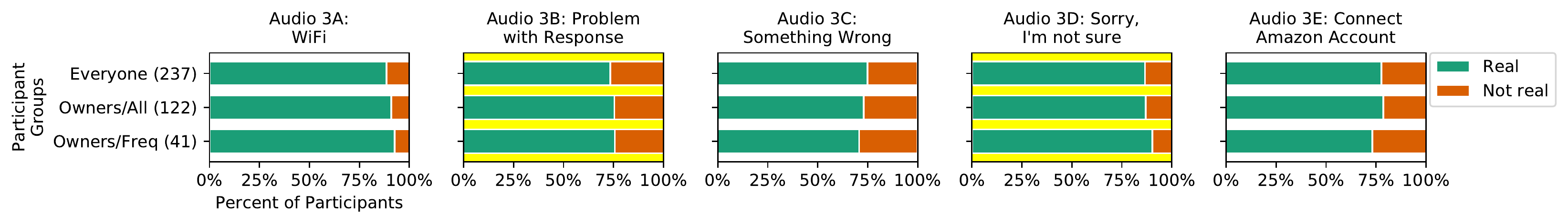}
\caption{Differentiating real system message from fake ones. Only the messages from Audios 3B and 3D are real (highlighted); the rest are fake.}
\label{fig:audio_3a_3e}
\end{figure*}

\subsection{Finding 2: Some participants cannot differentiate between native and third-party skills and messages}\label{sec:differentiate}

Even if users have an accurate Conceptual Model of Alexa with regard to
its third party skills, it is still crucial that they receive clear
Feedback~\cite{DesignOfEverydayThings} during conversations with Alexa
that suggest whether that exchange was with a third party.

In this section, we show that the majority of our participants were unable to
differentiate between native and third-party skills. A consequence of these
results is that Alexa users, even if they have an accurate Conceptual Model of
Alexa and its skills, might not get clear Feedback from Alexa with regard to
whether they have interacted with a third party. In fact, users might mistake
third party skills for native functionality, which can have serious
ramifications for their privacy and security.

\subsubsection{Differentiating between native and third-party skills}

In Survey Section 3, we asked participants to watch Videos 3A through 3E of a person
interacting with an Alexa device. After each video, we asked participants
whether the person in the video interacted with a native or third-party Alexa
skill. Only Video 3A referred to a native skill.

\Figure{video_3a_3e} presents the participant responses. Although the majority of participants could correctly identify Videos 3A and 3B (intended as easier examples), the accuracy was much lower for the remaining videos. In the worst case, only 3.0\% of ``Everyone'' and 4.9\% of ``Owners/Freq'' could correctly identify ``Parental control'' (Video 3E) as a third-party skill. This result shows that a user could potentially confuse a third-party skill -- whether malicious or not -- with what appears to be native functionality; the user may accidentally leak sensitive information to the unintended third party.

Additionally, experience and familiarity with Alexa did not always correlate with more correct responses. In fact, while 33.3\% of ``Everyone'' could correctly identify ``Baseball Scores'' (Video 3C) as a third-party skill, only 22.0\% of ``Owners/Freq'' could do so. This result is inline with our previous findings for Video 1B (\Figure{video_1a_1b}).

Participant responses after Videos 3A and 3C are particularly troubling. In
Video 3A, following the message's instructions (holding down the circle button
on an Echo device) would restart the system. In Video 3C, the fake message
prompts users to go to a website (in this case, just the Amazon website),
creating potential for a phishing attack if the website is not Amazon.com. The
fact that participants might accept system information verbally gives
potentially malicious skills significant leeway in the types of attacks they
might perform. For example, a fake skill could ask the user for their Wi-Fi
password or tell them their Alexa device is malfunctioning. A user's potential
inability to differentiate between real and fake system messages helps enable
voice squatting and masquerading attacks~\cite{zhang2019dangerous,
kumar2018skill}, and such attacks could be expanded to better incorporate system
messages (for example, a voice masquerading skill could respond with an error
message and then stay open).

\subsubsection{Differentiating between real (native) and fake messages (which we built)}

In Survey Section 3, we asked each participants to listen to Audios 3A through 3E. As shown in \Figure{audio_3a_3e}, a majority of participants
were unable to differentiate between real (i.e., as a result of native skills)
and fake (i.e., as a result of third-party skills) Alexa system messages. For example, 88.6\% of ``Everyone'' and 92.7\% of ``Owners/Freq'' thought that the ``WiFi'' message was real. Again, familiarity of Alexa may not be correlated with a higher rate of correct responses.

\begin{figure}[t]
\centering
\includegraphics[width=\columnwidth]{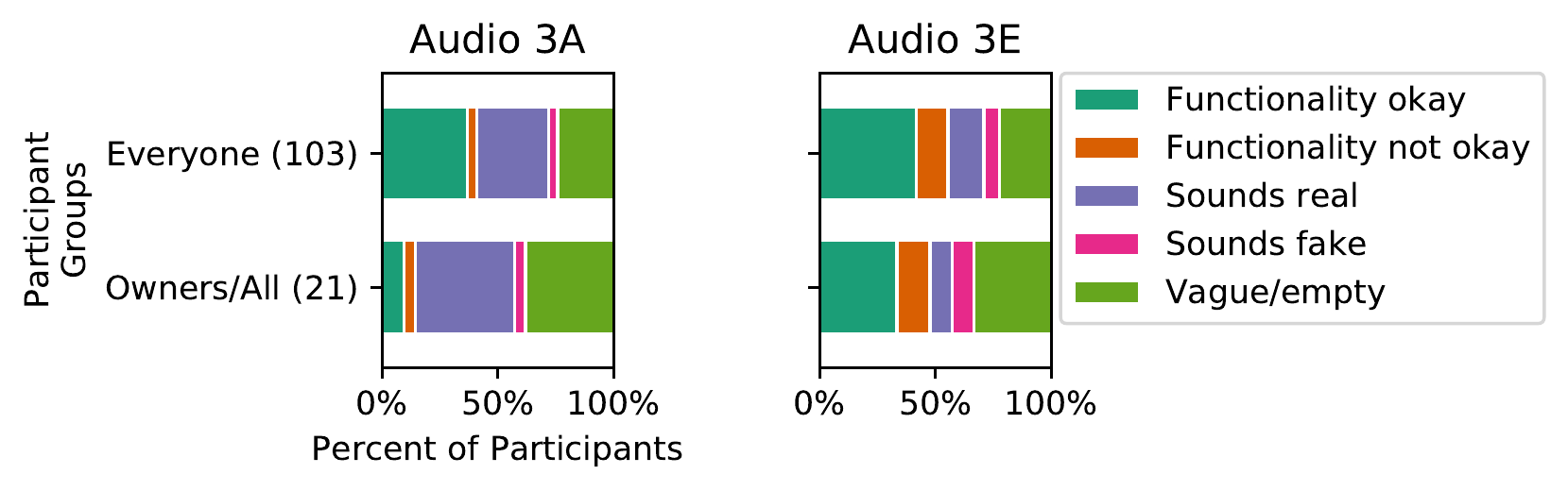}
\caption{User explanation (coded) for their responses to Audios 3A and 3E.}
\label{fig:audio_3a_3e_reason}
\end{figure}

After Audios 3A and 3E, we asked participants to briefly explain their answers in the University Survey.\footnote{We did not ask for free-text response in the MTurk survey to reduce the survey burden.} We grouped the responses into four categories: (i) functionality appearing to make sense (``Functionality okay''); (ii) functionality not making sense (``Functionality not okay''); (iii) audio sounding real or participant having heard it before (``Sounds real''); and (iv) audio sounding fake or participant never having heard it before (``Sounds fake''). We coded vague or empty responses as ``Vague/empty.''

We present the distribution of the codes in \Figure{audio_3a_3e_reason}. In both cases, ``Functionality okay'' and ``Sounds real'' dominate the reasons (while ignoring vague/empty responses). In particular, 33.3\% of ``Everyone'' thought Audio 3E's functionality made sense; for example, \textit{S1R23} said the response was real because \emph{``WiFi is necessary for Alexa function''}, and \textit{S1R42}, who responded it was fake because \emph{``I am not sure that Alexa has anything to do with Wifi.''} These responses suggest that some participants made judgements on the authenticity of a message based on whether the message was consistent with Alexa functionality. The fact that participants made these judgements based on functionality implies that a fake skill masquerading as the native system performing a reasonable task might seem believable; as will be presented later, participants did not have clear conceptions of what is reasonable on Alexa.

Furthermore, 9.6\% of ``Everyone'' felt the clip sounded real or claimed to have heard it before; for example, \textit{S1R8}, responded: \emph{``I've heard this one before''}, while \textit{S1R2} said the exact opposite: \emph{``I have not heard this previously.''} Similarly, \textit{S1R11} explained their response that the video was real based on Alexa's voice: \emph{``It sounds official?}''
These explanations suggest that participants made judgements based on the sound of a message. While these judgements are reasonable for users of a VUI, they can confuse users when Alexa uses the same voice for all functionality. This can be seen in the examples of participants that insisted they had heard Audio 3A before, which is impossible given we faked the message.

\subsection{Finding 3: Some participants do not know what voice commands
can invoke skills}\label{sec:understand}

Given previous findings that many users cannot differentiate between native and third-party skills, it is crucial that \textit{Discoverability} be well incorporated into Alexa's design~\cite{DesignOfEverydayThings}. If users are unable to differentiate third-party skills from native functionality, they need a clear understanding of Alexa standards with regard to third-party skills so that a third-party skill cannot mimic native functionality. In this section, we present results that suggest that users do not have a clear understanding of what phrases can invoke third-party Alexa skills and what verbal functionality the Alexa system does and does not provide.

\begin{figure*}[t]
\centering
\includegraphics[width=\textwidth]{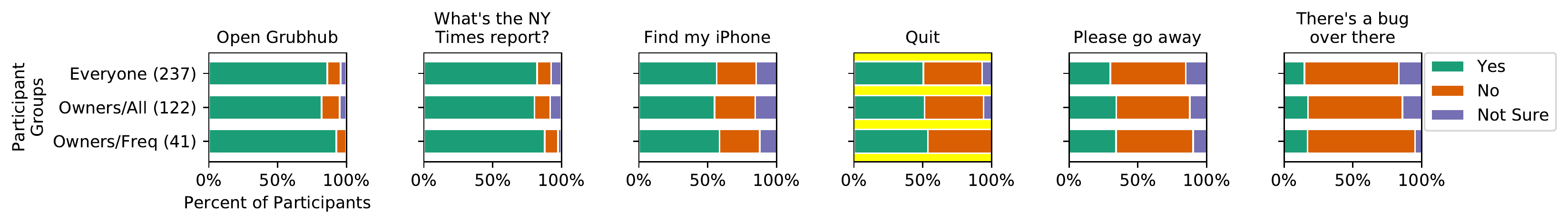}
\caption{Can each of the phrases above invoke an Alexa skill? In reality, all phrases, except ``Quit'' (highlighted), can invoke an actual skill.}
\label{fig:invocation}
\end{figure*}

\subsubsection{Users do not understand what phrases can invoke third-party skills on Alexa}

Many participants held incorrect assumptions regarding what phrases can invoke third-party Alexa skills. While many participants believed there were logical limits to what phrases can invoke an Alexa skill, in reality, nearly any  phrase is enough (as long as it begins with the wake word ``Alexa''). Although Amazon encourages developers to design skills with a few
recommended invocation phrases (such as ``Open <invocation name>'' and ``Ask
<invocation name> <some action>''), Alexa is designed to, at a minimum,
open skills by just their name~\cite{AVSSkill}. Since this name can be arbitrary, the invocation phrase is unbounded, thus creating challenges for Discoverability.

In Survey Section 4, we asked participants whether the six invocation phrases
could open skills on Alexa. As shown in \Figure{invocation}, most participants
understood that more conventional (based on Amazon's standards~\cite{AVSSkill})
phrases such as ``open Grubhub'' and ``what's the New York Times report?'' can
invoke skills on Alexa. In contrast, a majority of participants (54.8\% and
68.6\%, respectively) incorrectly responded that ``please go away'' and
``there's a bug over there'' cannot invoke skills on Alexa. Even though at the
time of writing no actual skills on the Amazon Store respond to such invocation
phrases, we successfully developed two private skills that could respond as such.

These results highlight a problem, especially given that users
often cannot differentiate native and third-party skills (as shown in Findings 2).
The fact that many
users may not understand which phrases can successfully invoke third-party
skills makes it even more likely they can accidentally invoke some skill and not
realize it has been built by a third party. It may also increase the
likelihood of invoking a malicious skill that can try to imitate the system or
mimic another skill ~\cite{zhang2019dangerous, kumar2018skill}. A salient example
of an attack that could leverage this result is the fake parental controls skill
presented in Finding 2, which a vast majority of participants believed was real
and native. Even if a malicious actor is not involved, users could still accidentally invoke a third-party skill without realizing so and transmit sensitive information to unintended third parties.

\begin{figure*}[t]
\centering
\includegraphics[width=\textwidth]{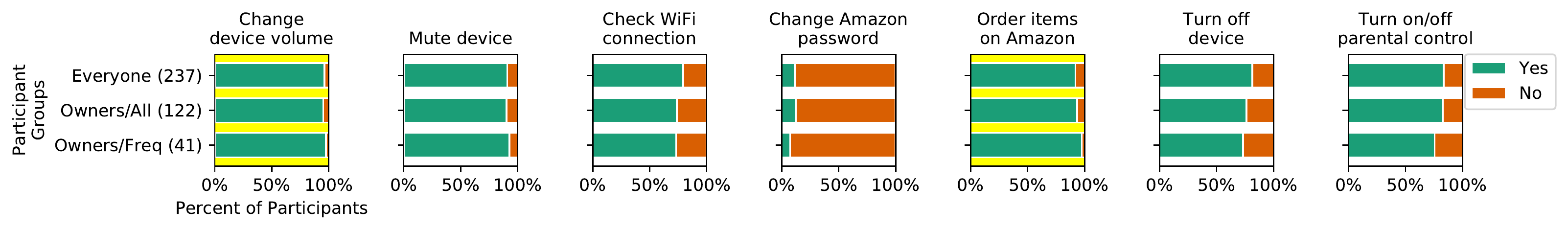}
\caption{Can you verbally do this with Alexa? In reality, users can only change the device volume and order items on Amazon (both highlighted) through verbal commands.}
\label{fig:functions}
\end{figure*}

\subsubsection{Some participants did not know what can and cannot be done with Alexa verbally}\label{sec:alexaverbally}

Participants often did not have clear intuitions regarding what can and cannot
be done with Alexa verbally (rather than through the app or with physical
buttons on the device).

As shown in \Figure{functions}, participants believed that most of the given tasks -- except changing the Amazon password -- could be done verbally with
Alexa, further expanding the potential attack space for malicious skills. In reality, only changing volume and ordering Amazon goods are feasible through Alexa's voice interface, although 90.7\% of ``Everyone'' thought they could verbally mute Alexa and 79.3\% of ``Everyone'' believed they could check the status of WiFi verbally. If a skill were to exist (whether malicious or benign) that responded to any of these invalid invocation phrases, a user may believe that he or she was interacting with the native system (especially given Findings 2) and potentially leak sensitive information.

It is worth pointing out that a vast majority (88.6\%) of ``Everyone''
did not believe one can verbally change their Amazon.com password with Alexa, presumably because changing password on non-verbal interfaces (e.g., on the web) could be the conventional practice and doing so over the VUI may deviate from this standard.
As such, there is potentially hope of raising
awareness for users to understand what can (e.g., changing volume) and cannot (e.g., changing passwords) be achieved natively on Alexa; this awareness would likely help users distinguish some third-party skills and native skills and protect their privacy.

\section{Recommendations for VUI design}

Some of Alexa's design decisions are inconsistent with Norman's design principles: Conceptual Model, Feedback, and Discoverability~\cite{DesignOfEverydayThings}. These inconsistencies likely led to the observations in our survey results. In this section, we propose design recommendations for Alexa -- and VUIs in general -- based on these  principles and our findings.

\subsection{Recommendation 1: Having clear indications of contexts}

Our results show that many participants were unable to distinguish between native and third-party skills (Finding 2), and this problem was compounded by the lack of awareness of third-party skills in the first place (Finding 1). These findings suggest that Alexa's design is inconsistent with the Conceptual Model and Feedback principles.

Our recommendation is for Alexa to clearly indicate the context to its users. This approach would provide users with the correct Conceptual Model that there are differences between native and third-party skills and among the third-party skills. Moreover, the approach would offer Feedback to users as to what context the interactions are in.

Past research in this realm has already yielded useful insights. To protect
against voice masquerading attacks, for instance,  Zhang et al. proposed a
``Skill Response Checker'' that checks VUI responses for phrases that can be
used to mimic the system~\cite{DBLP:journals/corr/abs-1805-01525}. Although such features
could be effective deterrents in some cases, our research
suggests that users might believe a wide array of messages, e.g., Audio 3A, 3C
and 3E, to be native system messages that would be difficult to blacklist
individually. Furthermore, our research suggests that privacy concerns can arise
even when skills are not trying to be malicious. Because users cannot always
differentiate between native and third-party skills, it is possible a
third-party skill might request information that, although not inherently
malicious, a user may not want to give.

One recommendation is for Alexa devices to show visual and audio cues. In
particular, Amazon Echo already uses the Light Ring to display system state,
such as powering on, listening, connecting to the network, or making phone
calls~\cite{alexa-light}. Alexa could indicate native and third-party contexts
using the Light Ring, e.g., flashing the lights as users switches from the native context to a third-party skill,
or showing different colors for native and third-party skills. Aside from visual
cues, Alexa could also leverage audio cues, e.g., using different voices for
native and third-party skills or playing a chime as a user switches from one
skill to another. The fact that 79.7\% of participants responded that Jeopardy
is a non-native skill (\Figure{video_3a_3e}) suggests that the change from Alexa
to Alex Trebek's voice may have tipped off users.

Although these recommendations may help a user distinguish between native and
third-party skills, there is a tradeoff between usability and transparency about
the origin of the skill. The visual cues are unlikely to be effective if users
do not look at the Light Ring (especially if an Amazon Echo device is stowed away
in a corner of a room and used primarily via voice). Also, there are already 12
distinct visual patterns on the Light Ring~\cite{alexa-light}; adding more
patterns to indicate context might further confuse users. The audio cues, on the
other hand, may be a distraction to users, as Amazon attempts to build a
seamless voice conversation experience where users are not expected to notice
the switch in the skill context~\cite{alexa-conversations}.

\subsection{Recommendation 2: Following consistent Alexa design standards}

Finding 3 shows that some participants do not know what commands Alexa can understand to invoke skills. This observation highlights a design of Alexa that is inconsistent with the Discoverability principle.

A comprehensive education of all available commands is unrealistic, as it
places unnecessary cognitive burden on the user. According to one
guide~\cite{cnn-voice-commands}, there are more than 200 commands to invoke
various native skills. Furthermore, for every new third-party skill invoked, a
user would have to remember the new commands associated with the skill.

Given that there are at least 47,000 skills available, Alexa could learn from
the Discoverability design principle~\cite{DesignOfEverydayThings} and follow common standards on what functions are and are not available on Alexa natively. For instance, it is possible for an Alexa user to change the volume but not mute the
device, set an alarm but not change the time zone, buy groceries but not music.
One simple solution is for all hardware-related commands to be strictly
non-verbal. Whenever the Alexa system detects a command for a hardware-related
feature such as changing the volume, it should clearly respond that such kinds
of commands cannot be done; currently, if a user asks Alexa to mute the device
or turn off, Alexa just ignores the command. Again, the exact design is not as
important as Amazon setting a consistent standard that it clearly shares with
developers and users.

Additionally, Alexa could impose strict standards on how to invoke skills. The fact that ``Please go away'' could actually invoke a skill (Finding 3) potentially threatens users' privacy. Although Amazon recommends certain common phrases for invoking
skills such ``Ask <invocation name> <some action>'', ``Tell <invocation name>
<some action>'', and ``Open <invocation name>'', any phrase (other than some
reserved for system functionality) can be used to open a third-party skill on
Alexa~\cite{AVSSkill}. This design creates a potentially confusing situation for
users. While many skills conform to common naming standards, Alexa's design leaves a backdoor for malicious skills to trick users or for one skill to accidentally obtain sensitive user information instead of the intended one. We recommend that Alexa follow a strict standard for invocation -- for instance, ``Open <invocation name>,'' but not any other phrases. Another recommendation is for Alexa to announce information about the skill, such as the developer's name, before running the skill for the first time; this approach could provide users with more transparency on the third party. However, these recommendations are, again, associated with usability trade-offs, because they make Alexa's VUI less flexible and more cumbersome to interact with and may go against Amazon's attempts to build a
seamless voice conversation experience~\cite{alexa-conversations}.

\section{Future Work}

Although most of our participants believed that the ``malicious'' skills we had
developed were native skills, it is unclear how often similar skills could be deployed on the Amazon Skill Store. If these malicious skills are prevalent, a user could confuse a malicious skill for a
native skill or another benign skill, or a user could confuse one benign skill
for another benign skill; in either case, the user could be revealing sensitive
information to an unintended third-party skill developer, which poses a privacy
risk.

To identify such skills in the wild, one of the challenges is scalability. In
particular, more than 47,000 skills are available on the Amazon Skill Store at
the time of writing. We would need to develop a method to programmatically
invoke each skill and, based on the skill's response, determine if the skill
resemble another native or third-party skill whether intentionally or
unintentionally, as such resemblance may cause confusion among users. This
automatic technique is difficult, because skills are executed remotely and each
verbal interaction with a skill is associated with an HTTP request~\cite{AVS1}.
Unlike mobile apps, for which we can use off-the-shelf-tools for static (e.g., by
downloading the binaries) and dynamic analyses (e.g., Android
Monkey~\cite{monkey}), we are not aware of any established tools or techniques
to analyze skills at scale. We defer this analysis to future work.

\section{Conclusion}

In this paper, we surveyed 237 new and existing users of Alexa devices. We found
that some participants were unaware that skills could be developed by third
parties, that most participants failed to distinguish native and third-party
skills and voice messages, and that they often did not understand what functions
or voice commands could be understood by Alexa. Surprisingly, participants with
more familiarity and experience with Alexa tended to show signs of confusion.
These findings suggest that a user may accidentally invoke an unintended skill
without being aware of this mistake; regardless of whether the skill is
malicious or benign, the uninteded third party may obtain sensitive user
information, thus giving rise to privacy risks. Our recommendations include
developing audio and visual indicators of native and third-party contexts, as
well as following a consistent design standard to help users learn what
functions are and are not possible on Alexa.

\bibliographystyle{IEEEtran}
\bibliography{\jobname}

\end{document}